# Research on fresh agricultural product based on the retailer's overconfidence under options and spot markets

**Kai Nie[1]**,
School of Economy and Trade, Hunan University, Changsha 410079, Hunan, China
d_niekai@163.com
**Man Yu[2]**
School of Economy and Trade, Hunan University, Changsha 410079, Hunan, China
lingsheshidubing@163.com
(1.School of Economy and Trade, Hunan University, Changsha 410079, Hunan, China

*ABSTRACT*: In this article, we analyze the application of options contract in the special commodity supply chain such as fresh agricultural products. This problem is discussed from the point of the retailer. When spot market and future market are both available, we discuss how the retailer chooses the optimal production. Furthermore, overconfidence is introduced to the supply chain of the fresh agricultural products, which has not happened before. Then，based on the overconfidence of the retailer, we explore how overconfidence affects the supply chain system under different circumstances. At last, we get the conclusion that different overconfidence level has different affection on retailer's optimal ordering quantity and profit.

## 1. Introduction

Agriculture is the basic department in one country's economy, supporting all human's life. Nearly all nations' governments pay more attention to the development of agriculture. In the No.1 document, it is proposed to speed up the development of the distribution channels of agriculture products, to improve the grades and levels of the circulation of them. It also emphasized that we need to make full use of the modern information technology, give full play to the agriculture future market to guide the production and averse the risk. The fresh agricultural products are the necessities in our daily life. It is very important for our health. With the development of our economy, we require more in freshness. Furthermore, developing fresh agriculture products is also an important method to enhance the income of the peasants. As one kind of special products, fresh agriculture products are featured with deteriorated and perished. Some may be wasted when circulating. In order to speed up the circulation of them and reduce wastage, some scholars research them in the point of supply chain. Cai[1] brought management of supply chain to the field of fresh agriculture, studied the coordination of fresh agricultural supply chain under the condition of pricing FOB. Chen Jun, Dan Bin etc. [2] introduced the freshness attenuation law to the agriculture product inventory model for deteriorating, which is expressed by the form of exponential function , explored the multi-level price discounts order policy of suppliers under the elasticity of demand condition. Wang lei and Li Yuyu [3] established the consumer utility model with considering the freshness, price and time variation, then, analyzed the retailer's optimal order strategy. Under the condition of the single three-stage supply chain, Lin Lue etc. [4] analyzed supply chain coordination of fresh agriculture product with time limited. Finally they gave the conclusion that the revenue sharing contract can be effectively coordinated.

With the deepening study of supply chain, more and more scholars introduce options into the supply chain. Barnes Schuster etc.[5] studied the application of options contracts in the two-stage supply chain system, in which the demand is correlated, reflecting the flexibility of options under uncertain conditions. Ritchken PH etc.[6] proposed that options contract is one form of contracts combining option with the traditional ordering methods. To a certain degree, it can improve the flexibility of supply chain, as well as strengthen cooperation in the supply chain. Chen Xiangfeng, Zhu Chenbo etc. [7] introduced options into the supply chain procurement, explained the process of adjusting the interests of all parties of the supply chain and risk-sharing by using option. Ning Zhong and Lin Bin [8] studied the application of stand-alone option mechanism and embedded options mechanism in supply chain management, analyzed the effects of options contract in supply chain performance, pointed out that the options could achieve the effect of revenue sharing and risk sharing. Many scholars also studied the supply chain of fresh agricultural products. Wang Jing etc. [9] drew options contract into fresh agricultural products, analyzed the wholesalers' ordering policies both in the spot market and future market at the same time, and obtained the optimal order quantities under the combination of both two kinds of markets. The defect of the article is ignoring the analysis of stock after introducing it.

Nowadays the wholesalers and retailers are more and more powerful than before, the dominance of the market has changed to some extent. However, some scholars' model designs lack rationality. Furthermore, former studies are based on the assumption of rational people, ignoring their behavioral characteristics and other irrational factors, such as overconfidence and so on. In actual transactions, the participants tend to be over-confident and more optimistic of their own abilities, knowledge, and predictions about the future performance[10]. Based on the above considerations, this paper attempts to establish a more reasonable model to analyze the optimal ordering policy of fresh agricultural product in the view of retailers.

## 2. Model Description

The model is explored by a retailer and a supplier of single-cycle supply chain system, circulating with fresh and perishable agricultural product. This paper is analyzed from the point of the retailer. The retailer can purchase the order both in the spot market and the option market. In the process of cooperation, the retailer decides the option price and the executed price [11], the supplier can ensure all orders of the retailer according to the contract. What's more, the retailer is overconfident of the market demand. In order to facilitate discussion, now the parameters of the model and their meanings are explained as follows:

w: wholesale price, $p$: sale price of fresh product, g: out of stock cost, x: the demand of product, $\mu$: expectation of market demand, $c_0$: option premium, $c_e$: executed price of option, $\beta$: the loss of fresh agricultural products in the unloading and transportation process, $0<\beta\leq 1$, $\theta$: freshness of products, $0<\theta\leq 1$, T: circle of ordering, $Q_1$: when the options contract is made by the wholesaler, the order quantity in the spot market by one cycle, $Q_q$: when the options contract is made by the supplier, the order quantity in the future market by one cycle, $F(x)$: the cumulative distribution function of fresh agricultural product market demand, $f(x)$: the probability density function of fresh agricultural product market demand, $F_r(x)$: the cumulative distribution function of the market demand of fresh agricultural product recognized by the wholesaler, $f_r(x)$: the probability density function of fresh agricultural product market demand recognized by the wholesaler.

Underlying assumptions are as follows:
(1) k presents the level of overconfidence of the retailer, $k\theta x$ is the demand of the market recognized by the retailer, when $0<k<1$, it tells the retailer's cognition of adverse signals of market demand, when $k>1$, it presents the retailer's cognition of favorable signals about the market demand.
(2) The initial inventory is zero, and the fresh agricultural products are not returnable, which is determined by their characteristics, and the remaining residual value is very small, they can be considered to be zero.
(3) The shortage cost is transferred to the supplier by virtue of the retailer's strong negotiation skills.
(4) $w_0 < c_e + c_0$, otherwise, retailers will order all the products in the options market.

## 3. Model analysis
### 3.1 The supplier's outcome policies

As the follower, the retailer determines its own ordering quantity to maximize its profit. Therefore, the retailer's expected profit can be expressed as:

$$E\pi_r = pE[\theta kx \wedge Q(1-\beta)] - c_0 Q_q(1-\beta) - c_e E[(\theta kx - Q_1(1-\beta)) \wedge Q_q(1-\beta)] - w_0 Q_1(1-\beta)$$
$$- gE[\theta kx - (Q_1 + Q_q)(1-\beta)]^+$$
$$= (p+g)(Q_1 + Q_q)(1-\beta) - (p+g-c_e)\theta k \int_0^{\frac{(Q_1+Q_q)(1-\beta)}{\theta k}} F(x)dx - (c_0 + c_e)Q_q(1-\beta) \quad (1)$$
$$- c_e k\theta \int_0^{\frac{Q_1(1-\beta)}{\theta k}} F(x)dx - w_0 Q_1(1-\beta) - g\theta k\mu$$

**Proposition 1: When the supplier supports the shortage cost, its optimal outcome of fresh agriculture product is**

$$Q^* = \frac{k\theta}{1-\beta} F^{-1}(\frac{p+g-c_e-c_0}{p+g-c_e}), \quad Q_1^* = \frac{\theta k}{1-\beta} F^{-1}(\frac{c_0+c_e-w_0}{c_e}), \quad Q_q^* = Q^* - Q_1^* \quad (2)$$

Proof: We take a derivative with respect to $Q_q$、$Q_1$, there is

$$\frac{\partial E\pi_r}{\partial Q_q} = (p+g)(1-\beta) - (p+g-c_e)(1-\beta)F(\frac{1}{\theta k}(Q_1+Q_q)(1-\beta)) - (c_0+c_e)(1-\beta)$$

$$\frac{\partial E\pi_r}{\partial Q_1} = (p+g)(1-\beta) - (p+g-c_e)(1-\beta)F(\frac{1}{\theta k}(Q_1+Q_q)(1-\beta))$$
$$- c_e(1-\beta)F(\frac{1}{\theta k}Q_1(1-\beta)) - w_0(1-\beta)$$

Order $\frac{\partial E\pi_r}{\partial Q_q} = 0$, $\frac{\partial E\pi_r}{\partial Q_1} = 0$, then, we can get $Q^* = \frac{k\theta}{1-\beta} F^{-1}(\frac{p+g-c_e-c_0}{p+g-c_e})$,

$Q_1^* = \frac{\theta k}{1-\beta} F^{-1}(\frac{c_0+c_e-w_0}{c_e})$, here, Proposition was proofed.

Put (2) in (1), the retailer's optimal expected profit can be written as follows:

$$E\pi_r^* = (p+g)Q^*(1-\beta) - (p+g-c_e)\theta k \int_0^{\frac{Q^*(1-\beta)}{\theta k}} F(x)dx - (c_0+c_e)Q_q^*(1-\beta)$$
$$- c_e\theta k \int_0^{\frac{Q_1^*(1-\beta)}{\theta k}} F(x)dx - w_0 Q_1^*(1-\beta) - g\theta k\mu \quad (3)$$

Under the condition of decentralized decision making, the supplier's profit is:

$$\pi_s = w_0 Q_1(1-\beta) + c_0 Q_q(1-\beta) + c_e \min[\theta x - Q_1(1-\beta), Q_q(1-\beta)] - cQ, \quad (4)$$

Then, its expected profit is

$$E\pi_s = w_0 Q_1^*(1-\beta) + (c_0+c_e)Q_q^*(1-\beta) - c_e\theta \int_0^{\frac{Q^*(1-\beta)}{\theta}} F(x)dx + c_e\theta \int_0^{\frac{Q_1^*(1-\beta)}{\theta}} F(x)dx - cQ^* \quad (5)$$

### 3.2 Analysis of the impact of overconfidence
#### 3.2.1 Impact of Overconfidence on retailer's ordering quantity

**Proposition2: the optimal total ordering quantity and spot market quantity are increasing function of overconfidence level.**

When the retailer is entirely rational, it can judge the real demand of the market, according to the real demand, it decides its ordering quantity. So, the totally rational retailer's profit is as follows:

$$E\pi_{r0} = (p+g)(Q_1+Q_q)(1-\beta) - (p+g-c_e)\theta \int_0^{\frac{(Q_1+Q_q)(1-\beta)}{\theta}} F(x)dx \qquad (6)$$

$$-(c_0+c_e)Q_q(1-\beta) - c_e\theta \int_0^{\frac{Q_1(1-\beta)}{\theta}} F(x)dx - w_0 Q_1(1-\beta) - g\theta\mu$$

Then, we can get the totally rational retailer's optimal ordering quantities:

$$Q_0^* = \frac{\theta}{1-\beta} F^{-1}(\frac{p+g-c_e-c_0}{p+g-c_e}), \quad Q_{01}^* = \frac{\theta}{1-\beta} F^{-1}(\frac{c_0+c_e-w_0}{c_e}), \quad Q_{0q}^* = Q^* - Q_1^* \qquad (7)$$

Compared the expression (2) to (7), we know that the optimal total ordering quantity and spot market ordering quantity are both linear functions of the level of overconfidence k, there is a positive correlation between k and optimal total ordering quantity. It can be expressed as follows: When a favorable signal is shown on the market, it means k> 1, the higher level of overconfidence of the retailer is, the greater $Q^*$ and $Q_1^*$ are; When there are adverse market signals, that is to say 0< k <1, more overconfident the retailer is, smaller $Q^*$ and $Q_1^*$ are. Option ordering quantity has no relationship with overconfidence coefficient.

**(2). Impact of Overconfidence on supplier's profit**

From the optimal ordering quantities and optimal outcome, we could get

$$E\pi_{s0} = w_0 Q_1(1-\beta) + (c_0+c_e)Q_q(1-\beta) - c_e\theta\int_0^{\frac{Q^*(1-\beta)}{\theta}} F(x)dx + c_e\theta\int_0^{\frac{Q_1^*(1-\beta)}{\theta}} F(x)dx - cQ \qquad (8)$$

**Proposition 3: Because of the overconfidence of the retailer, the profit of supplier is not stable. That is to say its profit is very easy to change.**

Proof: compared the two kinds of profit, we know

$$E\pi_{s0} - E\pi_s = (1-k)[w_0 Q_1(1-\beta) + (c_0+c_e)Q_q(1-\beta) - c_e\theta\int_0^{\frac{Q^*(1-\beta)}{\theta k}} F(x)dx \qquad (9)$$

$$+ c_e\theta\int_0^{\frac{Q_1^*(1-\beta)}{\theta k}} F(x)dx - cQ^*]$$

When 0<k<1, $E\pi_{s0} - E\pi_{s1} > 0$, the retailer's overconfidence will reduce the supplier's profit.

What's more, the higher the overconfidence level is, the smaller the supplier's profit is. When k>1, $E\pi_{s0} - E\pi_{s1} < 0$, the retailer's overconfidence will improve the supplier's profit.

### 3.3 The optimal decisions under Centralized decision-making condition

Under centralized decision-making, we aims to research on the whole supply chain performance status. At this point, as the decision maker, the supplier treats maximizing profit of the whole supply chain as the goal, to decide the optimal production. By assumes, we know that the supplier is fully rational, so the whole supply chain profit can be expressed as:

$$E\pi = pE[\theta x \wedge Q(1-\beta)] - cQ - gE[\theta x - Q(1-\beta)] \qquad (10)$$

$$= (p+g)Q(1-\beta) - (p+g)\theta\int_0^{\frac{Q(1-\beta)}{\theta}} F(x)dx - cQ - g\theta\mu$$

Take derivative with respect to $Q_q$、$Q_1$,

$$\frac{\partial E\pi}{\partial Q} = (p+g)(1-\beta) - (p+g)(1-\beta)F(Q(1-\beta)/\theta) - c,$$ then, we could get the optimal ordering

quantity $Q^{**} = \frac{\theta}{1-\beta} F^{-1}(\frac{(p+g)(1-\beta)-c}{(p+g)(1-\beta)})$. （11）

### 3.4 Control and adjustment of retailer's overconfidence level

When the retailer is overconfident about the market demand of fresh agricultural product, the interests of both sides will change with the level of retailer's overconfidence. At this time, the supply chain as a whole does not reach the best situation. What's more, the fluctuation in the level of overconfidence is not conducive to the cooperation between the two sides. It is necessary to find a suitable way to eliminate the impact of retailer's overconfidence. Croson [12] designed a buy-back contract and wholesale-price contract to eliminate the retailer's overconfidence in general supply chain. Due to the higher requirements of fresh agricultural product, the perishable feature determines the products cannot be returned. As a flexible tool, options can be used to achieve the coordination of the supply chain. In this article, we attempt to use the option contract to control the retailer's overconfidence, so as to coordinate the fresh agriculture supply chain at last.

**Proposition 4: when $c_0$ and $c_e$ satisfy $c_0 = (p+g-c_e)[1-F(\frac{1}{k}F^{-1}(\frac{(p+g)(1-\beta)-c}{(p+g)(1-\beta)}))]$, the outcome is optimal.**

Proof: from the context we know that the profit of the whole supply chain can be maximized just while $Q^{**} = Q^*$, $\frac{\theta}{1-\beta} F^{-1}(\frac{(p+g)(1-\beta)-c}{(p+g)(1-\beta)}) = \frac{\theta k}{1-\beta} F^{-1}(\frac{p+g-c_e-c_0}{p+g-c_e})$, solve this equation, we get $c_0 = (p+g-c_e)[1-F(\frac{1}{k}F^{-1}(\frac{(p+g)(1-\beta)-c}{(p+g)(1-\beta)}))]$, by now, the proposition is proofed.

Furthermore, according to (12), the anti correlation exists in option price and the overconfidence coefficient, the higher the retailers' overconfidence level is, the smaller the option price.

## 4. Numerical example

Her we use matlab7.0 to do the numerical analysis. Assume there is a supermarket ordering fresh fish from a supplier. They can finish the trade both in the spot market or the option market. Here we just analyze the profit of the supermarket. We assume that the market demand is subject to normal distribution. Wholesale price $w_0 = 25$, sale price $p = 50$, out of stock cost $g=10$, $\beta = 0.1$, $\theta = 0.8$, the fresh agriculture product demand obeys uniform distribution between (0,100), the initial option price $c_0 = 5$, then $c_e = 60 - \frac{90k}{18k-13}$, assume $c_e = 35$, then $c_0 = 25 - \frac{325}{18k}$. When other parameters are established, the optimal order quantity $Q$ and market order $Q_1$ can be calculated. By calculating, the relationship between optimal order quantities and the overconfident coefficient can be described as follows:: $Q^* = 64.2$, $Q_1^* = \frac{800k}{9}[1 - \frac{20(18k-13)}{60(18k-13)-90k}]$, $Q_q^* = Q^* - Q_1^*$, the relationship between optimal profit of the retailer and the overconfident coefficient:

$E\pi_{s1} = \frac{1587600k^4 - 57960000k^3 + 8738701k^2 - 6048302k + 1561560}{3(18k-13)(33k-26)}$。

The relationship among the variables is reflected in graphs as follows:

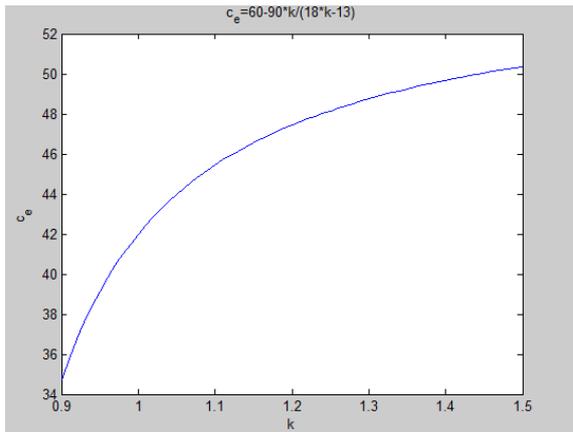
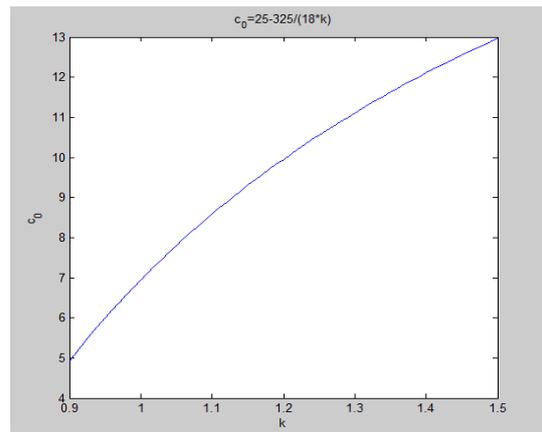

Pic1: the relationship between $c_e$ and k

Pic2: the relationship between $c_0$ and k

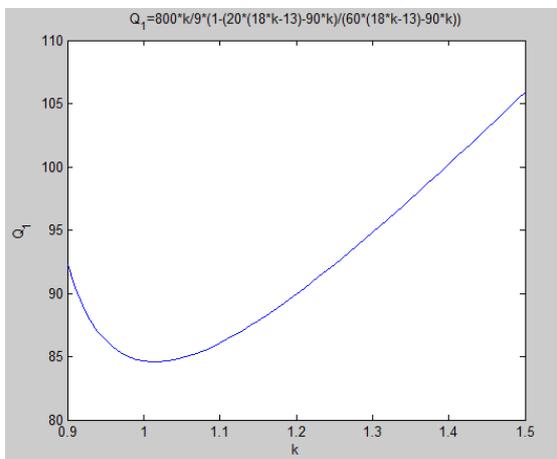
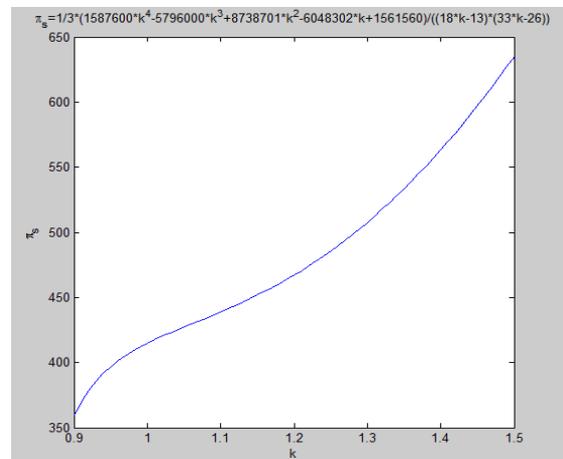

Pic3: the relationship between $Q_1^*$ and k

Pic4: the relationship between $\pi_s$ and k

Figure 1 and figure 2 show that the option price and the executed price of option are increasing functions of overconfidence level k, they have the same changing direction of overconfidence coefficient. Figure 3 shows that on the premise of realizing supply chain coordination, when k = 1, namely when the retailer is perfectly rational, its spot order is minimum; When the retailer's overconfidence exists, the physical quantity is greater than rational, and the higher the retailer's overconfidence level is, the larger its spot ordering quantity is. Figure 4 shows that when the market appears negative signals, the more rational the retailer is, the greater the profits of the supplier is; and when the market is in a good indication, the higher the retailer's overconfidence level, the greater the profits of supplier. By analysis of the above situation, we know it is consistent with theoretical model and actual situation.

## 5. Conclusion

This paper changes the mode of the traditional supply chain that the retailer is fully rational, in fact, the irrational factors are always exiting, such as overconfidence. In this article we analyzed the impact of overconfidence on the fresh agriculture product supply chain. Furthermore, based on the influence of the retailer's overconfidence on the prediction of market demand, we analyzed the case of the wholesaler setting options, how the overconfident coefficient affects the optimal quantities and the profits as well as how the optimal ordering quantity varies. Finally, we could draw the conclusion that the optimal order quantities are related to the level of overconfidence. When the prospect of market is good, overconfidence of the retailer has a positive effect, and vise verse. Nowadays, with the buyers' status and role gradually improving, the initiative also correspondingly increases in the market. By the way, they will develop more options contracts which are conducive to themselves. This article just analyses the single-cycle, one-to-one mode of supply chain, so more detailed and deeper research could be made about order strategy of fresh agricultural products.

**Acknowledgement:** This article is supported by Province natural science fund project (Modeling and optimizing the cross-regional emergency logistics information integration system 13JJB001). Thanks for its help and support.